\documentclass[lettersize,journal]{IEEEtran}
\usepackage{amsmath,amsfonts}
\usepackage{array}
\usepackage[caption=false,font=normalsize,labelfont=sf,textfont=sf]{subfig}
\usepackage{textcomp}
\usepackage{stfloats}
\usepackage{url}
\usepackage{verbatim}
\usepackage{graphicx}
\usepackage{cite}
\usepackage{amsmath,amssymb,amsfonts}
\usepackage{multirow}
\usepackage{rotating}
\usepackage{booktabs}
\usepackage{balance}
\usepackage{acronym}
\usepackage{tabularx}
\usepackage{booktabs}

\usepackage{algorithm}
\usepackage[noend]{algpseudocode}

\newcommand{\refeq}[1]{(\ref{#1})}
\newcommand{\reffig}[1]{Figure~\ref{#1}}

\newcommand{\refalg}[1]{Algorithm~\ref{#1}}
\newcommand{\reftab}[1]{Table~\ref{#1}}

\acrodef{RL}{Reinforcement Learning}
\acrodef{RoA}{Region of Attraction}
\acrodef{MDP}{Markov desision process}
\acrodef{PPO}{Proximal Policy Optimization}
\acrodef{TRPO}{Trust Region Policy Optimization}
\acrodef{SDP}{Semi-Definite Program}
\acrodef{SOS}{Sum-of-Squares}
\acrodef{KL}{Kullback-Leibler}
\acrodef{TPC}{Tree-Structured PPO Controller}
\acrodef{GPC}{Graph-Structured PPO Controller}

\begin{document}

\title{Deep Reinforcement Learning Graphs: Feedback Motion Planning via Neural Lyapunov Verification}

\author{Armin~Ghanbarzadeh,~\IEEEmembership{Member,~IEEE}, Esmaeil~Najafi,~\IEEEmembership{Member,~IEEE}
}

%

\maketitle

\begin{abstract}
Recent advancements in model-free deep reinforcement learning have enabled efficient agent training. However, challenges arise when determining the region of attraction for these controllers, especially if the region does not fully cover the desired area. This paper addresses this issue by introducing a feedback motion control algorithm that utilizes data-driven techniques and neural networks. The algorithm constructs a graph of connected reinforcement-learning based controllers, each with its own defined region of attraction. This incremental approach effectively covers a bounded region of interest, creating a trajectory of interconnected nodes that guide the system from an initial state to the goal. Two approaches are presented for connecting nodes within the algorithm. The first is a tree-structured method, facilitating "point-to-point" control by constructing a tree connecting the initial state to the goal state. The second is a graph-structured method, enabling "space-to-space" control by building a graph within a bounded region. This approach allows for control from arbitrary initial and goal states. The proposed method's performance is evaluated on a first-order dynamic system, considering scenarios both with and without obstacles. The results demonstrate the effectiveness of the proposed algorithm in achieving the desired control objectives.
\end{abstract}


\section{Introduction}
\label{sec:introduction}

\IEEEPARstart{T}{his} study addresses the challenge of controlling underactuated nonlinear dynamic systems characterized by input saturation and bounded states. Existing methodologies involve the linearization of such systems and the subsequent design of controllers, albeit with the limitation that the resultant controller is only effective in the vicinity of the system's origin. In contrast, data-driven approaches, particularly those leveraging \ac{RL}, offer a promising avenue for addressing this limitation. \ac{RL}, employing deep function approximators, can effectively design controllers tailored to highly nonlinear systems. However, it is essential to acknowledge certain drawbacks associated with these algorithms. Firstly, they entail a substantial computational cost, posing challenges in real-time applications. Furthermore, a critical limitation lies in the absence of explicit confirmation of stability following the computation of an \ac{RL} controller. Consequently, users are left with an inherent uncertainty regarding the viability of the controller within a specific region, commonly referred to as the \ac{RoA}. This lack of information regarding the controller's effective operational domain constitutes a noteworthy concern in the practical application of \ac{RL}-based control strategies.

\begin{figure}[t]
\centering
\subfloat{\includegraphics[width=0.45\linewidth]{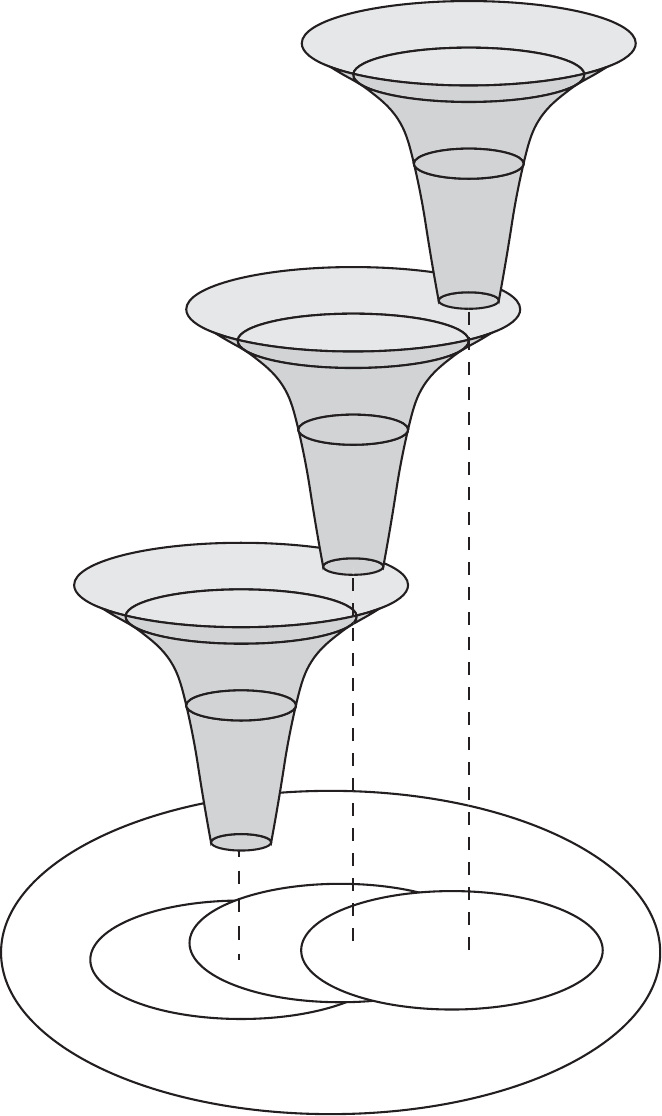}}
\caption{Illustration of the sequential controllers path following akin to a ball descending through funnels. Each controller is only active outside the domains of lower controllers. The lowest controller stabilizes the system at the final destination. Each funnels preimage represents the \ac{RoA} of the corresponding controller, adopted from \cite{burridge1999sequential}}
\label{fig:funnel}
\end{figure}

In order to facilitate the applicability of linear controllers to nonlinear systems, the concept of sequential control has been introduced, as outlined in \cite{burridge1999sequential}. This approach involves the initial design of linear controllers without explicit consideration of their \ac{RoA}. Subsequently, each controller undergoes evaluation to ascertain its respective \ac{RoA}. The subsequent controller is then designed with the deliberate intention of having its \ac{RoA} situated within the preceding domain, while concurrently stabilizing a distinct point within the operational state space of the system. This iterative process, referred to as synthesis, involves the continued design of these local controllers, and by following the trajectory they define, the system can progress from an initial state to a desired state. The path-following mechanism employed by these sequential controllers can be analogously likened to a ball descending through sequentially arranged funnels, as illustrated in Figure \ref{fig:funnel}. Each funnel's mouth, or preimage, corresponds to the \ac{RoA} of the associated controller. Upon traversing through a funnel, the ball transitions to the next one, and this process repeats until the system reaches its intended goal \cite{1087242}.

Various algorithms have been developed to address the challenges of efficiently and safely planning trajectories for robots. Classical algorithms such as A* \cite{hart1968formal} and Dijkstra's algorithm \cite{DIJKSTRA1959} have laid the foundation, but recent advancements have introduced a plethora of novel approaches.

Randomized motion planning methods have emerged as powerful tools in the field of robotics and autonomous systems, offering efficient solutions to complex path planning problems \cite{kavraki1994randomized, barraquand1993nonholonomic}. These methods, often rooted in probabilistic algorithms, introduce an element of randomness to the planning process, enabling robots to navigate through dynamic and uncertain environments. Notable among these techniques is the Rapidly-exploring Random Tree (RRT) \cite{lavalle2001rapidly} algorithm which has the ability to efficiently explore configuration spaces by randomly sampling feasible paths. Additionally, variants such as RRT* \cite{karaman2010incremental}, RRT*-Smart \cite{islam2012rrt}, and A*-RRT \cite{brunner2013hierarchical} have further improved upon the original framework, addressing issues related to optimality and computational efficiency. LQR-RRT a was proposed as a kinodynamic variant for complex or underactuated dynamics \cite{perez2012lqr}. Fast Marching Trees (FMT) \cite{janson2015fast} is a variant of the RRT algorithm designed to efficiently explore and plan paths in high-dimensional configuration spaces. FMT combines ideas from the Fast Marching Method with the probabilistic nature of RRT. The algorithm iteratively grows a tree from the starting configuration, favoring regions with lower cost-to-go values. FMT is particularly well-suited for applications in motion planning for robotic systems operating in dynamic and challenging environments. Probabilistic Roadmap (PRM) \cite{kavraki1996probabilistic}, a widely employed randomized motion planning method in robotics, constructs a graph representation of the configuration space through random sampling and collision checking. PRM has various variants, including PRM* \cite{ladd2004measure}, which enhances optimality by refining the roadmap over iterations demonstrating adaptability and efficiency in navigating complex and high-dimensional environments.

In the field of controlling dynamic systems, a common issue that arises after trajectory planning is that designing controllers for these paths can be tricky and less effective compared to other trajectory options. A more recent solution involves using feedback motion planning algorithms, which tackle both path planning and controller design simultaneously. \cite{atkeson2007random} used a combination of local trajectory optimizers and a stabilizing technique called Linear Quadratic Regulator (LQR) with random starting points. The method proposed exploring the solution space thoroughly in hops of finding the best overall solution. \cite{goebel2009smooth} suggested a control design approach using "patchy Lyapunov functions." This method stabilizes the system by strategically covering the state space with a set of locally effective control functions and ensuring stability through a switching feedback control.

The method proposed in this work is motivated from the \ac{RoA}-based planning method proposed in \cite{tedrake2010lqr}, in which the authors propose a feedback motion-planning algorithm that leverages rigorously computed stability regions to construct a sparse tree of LQR-stabilized trajectories. The resulting non-linear feedback policy exhibits a region of attraction that "probabilistically covers" the entire controllable subset of the state space. This paper endeavors to integrate advancements in data-driven methodologies for the control of nonlinear dynamic systems, with a particular focus on the application of \ac{RL} to formulate local controllers within the state space. Utilizing neural Lyapunov networks, the determination of the \ac{RoA} for these controllers is facilitated, thereby achieving stabilization of the local controller. The principal contributions of this study encompass:

\begin{enumerate}
\item Implementation of model-free data-driven controllers for the stabilization of dynamic systems, executed in a sequential and localized fashion.
\item Validation of the attraction boundary of sequential controllers based on the Lyapunov stability criterion, employing Neural Networks as Lyapunov function approximators derived from empirical data obtained from the system.
\item Introduction of a tree-based methodology that constructs a hierarchical tree of controllers to navigate from an initial state to a goal state.
\item Introduction of a graph-based approach that establishes an interconnected graph of controllers, enabling the traversal from any initial state to a goal state within a bounded region of interest.
\item Presentation of methodologies for integrating obstacle avoidance strategies within the design of controller sequences.
\end{enumerate}

The paper is structured as follows: In Section \ref{sec:background}, foundational principles of deep \ac{RL} controllers are introduced, along with background on Lyapunov stability theory and steps for designing neural networks to approximate Lyapunov functions. Section \ref{sec:controller} establishes the groundwork for the proposed algorithm, highlighting the pivotal role of nodes in deep \ac{RL}-based controllers and explaining the methodology for training and stabilizing these controllers. Two approaches for interconnecting nodes, forming either a tree (Section \ref{sec:tree}) or a graph (Section \ref{sec:graph}) of interconnected controllers, are presented, enabling the tracing of a trajectory to attain a goal state, with details of implementation and evaluation of the algorithms on a two-dimensional input-saturated dynamic system to demonstrate their effectiveness. Section \ref{sec:discussion} addresses addresses the main differences of each approach, and and provides a comparison of both methods advantages and trade-offs. Finally, Section \ref{sec:conclusion} concludes with a reflective summary of the research contributions.

\section{Background}
\label{sec:background}

Since the controllers synthesized in this work are Data-driven \ac{RL} based controllers, the need for neural based Lyapunov functions arise. When paired together, this gives us a powerful tool to control and calculate the applicable region without the need of explicit system dynamics, thus enabling us to chain the controllers in sequence. 

\subsection{Deep Reinforcement Learning}

In the domain of control and reinforcement learning, 

an agent is tasked with the challenge of selecting a sequence of actions to maximize the cumulative reward. This challenge is formalized as a \ac{MDP} \cite{sutton2018reinforcement}, defined by the tuple $(S, A, F, R)$. Here, $S$ represents the state space, $A$ denotes the action space, $R: S \times A \to \mathbb{R}$ is the reward function, and $F: S \times A \to S$ describes the state transition function, revealing the dynamics of the system. A policy, denoted as $\pi_\theta: S \to P(A)$, is employed to make action selections at each state within the \ac{MDP}. Typically stochastic, the policy is characterized by $\theta \in \mathbb{R}^n$, where $\theta$ is a vector encompassing $n$ parameters. The agent employs this policy to interact with the \ac{MDP}, resulting in a trajectory of states, actions, and rewards denoted as $\tau_{1:T} = s_0, a_0, r_0, \ldots, s_T, a_T, r_T: S \times A \times \mathbb{R}$. The primary goal of the agent is to compute a policy that maximizes the cumulative discounted reward across the trajectory obtained through policy sampling. This goal is conceived as the performance objective
\begin{equation}
J(\pi_\theta) = \mathop \mathbb{E} \limits_{\tau \sim \pi_\theta} \left[ {\sum\limits_{t = 0}^T {{\gamma ^t}R_t}} \right],
\end{equation}
where $\gamma$ is a discount factor often set close to 1. Policy methods, a subset of \ac{RL} algorithms, are specifically crafted to directly obtain the optimal target policy $\pi^*$. This approach differs from value-based methods \cite{mnih2015human, lillicrap2015continuous}, which first require the estimation of value functions. While value-based methods focus on approximating the value of states or state-action pairs, policy methods directly aim to find the optimal policy that dictates the agent's actions in various states. This distinction in approach reflects a fundamental divergence in the strategies employed by these two classes of \ac{RL} algorithms. The distinctive feature of policy-based methods lies in their capacity for direct optimization through gradient ascent techniques, involving the calculation of the gradient of the performance metric concerning the policy parameters \cite{sutton1999policy}
\begin{equation}
{\nabla _\theta }J({\pi _\theta }) = \mathop \mathbb{E} \limits_{\tau \sim \pi_\theta} \left[ {\sum\limits_{t = 0}^T {\nabla _\theta  } \log \pi_\theta (a_t | s_t) {{\hat A}_t}} \right],
\end{equation}
where ${\hat A}_t = {\hat Q}_t - {\hat V}_t$ is an estimator of the advantage function at timestep $t$. 

Within policy gradient methods, policy update involves making incremental adjustments in the parameter space to optimize the objective function. Nonetheless, modest changes in the parameter space may yield substantial alterations in the policy distribution, potentially leading to worse results. To address this challenge and ensure convergence, it becomes necessary to intermittently constrain the adjustments in the parameter space. The \ac{TRPO} algorithm \cite{schulman2015trust} addresses this concern by introducing a penalty term based on the \ac{KL} divergence constraint between the updated and the previous policy. Another approach, implemented in the \ac{PPO} \cite{schulman2017proximal}, maintains the advantages of \ac{TRPO} while proposing a more straightforward implementation through a revised objective function. The "Clipped Surrogate Objective" used in \ac{PPO} is
\begin{equation}
L^{CLIP}(\theta) = \mathop \mathbb{E} \limits_{\tau \sim \pi_\theta}
 \left[ \min (r_t(\theta){\hat A}_t, \mathrm{clip} (r_t(\theta), 1-\epsilon, 1+\epsilon) {\hat A}_t) \right],
\end{equation}
where $r_t{\theta} = \frac{\pi_\theta (a_t | s_t)}{{\pi_\theta}_\mathrm{old} (a_t | s_t)}$ is the probability ratio of an policy and $\epsilon$ is a hyperparamter (e.g. $\epsilon = 0.2$) that sets the bounds of the probability ratio. The employed objective function systematically applies parameter update clipping, selectively restraining updates only when they would degrade the probability and disregarding changes that result in an enhancement of the objective function.

In many practical decision-making problems, the states $s$ within the \ac{MDP} are characterized by high dimensionality, such as images captured by a camera or raw sensor data from a robot, rendering conventional \ac{RL} algorithms ineffective. Deep \ac{RL} algorithms integrate deep learning techniques to address such complex \ac{MDP}s, often representing the policy $\pi (a|s)$ or value functions through neural networks. These approaches have demonstrated remarkable efficacy, outperforming alternative methods across various tasks and achieving performance levels comparable to or surpassing those of professional human operators \cite{mnih2013playing}. In this study, the \ac{PPO} algorithm has been employed, utilizing deep neural networks for both policy and value network approximations. \ac{PPO} is a popular model-free on-policy policy gradient method which is simpler to implement compared to similar performing algorithms, still being sample-efficient, showing robust convergence, generalizing to different tasks and environments as well as being parallelizable and fast with regards to the wall-clock time \cite{gao2020deep, vinyals2019grandmaster, yu2022surprising}. Additional details on the network architecture are provided in Section \ref{sec:controller}.

\subsection{Lyapunov Neural Network}

A dynamic system of dimensionality $n$ is described by the differential equation $\dot{x} = f_u(x)$, with an initial condition of $x(0) = x_0$. Here, $f_u: \mathcal{D} \to \mathbb{R}^n$ represents a continuous vector field, where $\mathcal{D} \in \mathbb{R}^n$ is an open set containing the origin, defining the system's state space. The feedback control is governed by a continuous time-invariant function $u: \mathbb{R}^n \to \mathbb{R}^m$, which is incorporated into the full dynamics expressed by $f_u$. According to the stability criteria outlined in \cite{khalil2009lyapunov}, an equilibrium point $x^*$ is asymptotically stable if there exists a function $V: \mathbb{R}^n \to \mathbb{R}$ that satisfies the following conditions:
\begin{equation}
\label{eq:lyapunov_criterion}
V(x^*) = 0;~\mathnormal{and}~\forall x \in \mathcal{D} \setminus {x^*}, V(x)>0~\mathnormal{and}~\nabla_{f_u} V(x)<0,
\end{equation}
then $V$ is called a Lyapunov function. The Lie derivative of a continuously differentiable function $V$ over a vector field $f_u$ is defined as 
\begin{equation}
\nabla_{f_u} V(x) = \sum\limits_{i = 1}^n {\frac{{\partial V}}{{\partial {x_i}}}} \frac{{d{x_i}}}{{dt}}.
\end{equation}
The challenge in Lyapunov stability analysis resides in the identification of a suitable Lyapunov function for a given system. This function must adhere to the specific conditions to guarantee the stability of the system.

Various computational methods for identifying Lyapunov functions are surveyed in \cite{giesl2015review}. When the system dynamics are polynomial or can be effectively approximated by polynomials, an efficient approach involves computing a Lyapunov function through \ac{SDP} \cite{boyd2004convex}. The Lyapunov function can be further constrained to a \ac{SOS} polynomial \cite{parrilo2000structured}. Other methodologies for computing \ac{RoA}s encompass the maximization of a measure of \ac{RoA} volume along system trajectories \cite{henrion2013convex}. Additionally, there are sampling-based techniques that extend stability information from discrete points to continuous regions \cite{bobiti2016sampling, najafi2016fast}.

Recent research has concentrated on leveraging neural networks to acquire Lyapunov functions for controllers. In \cite{richards2018lyapunov}, the emphasis is on discrete-time polynomial systems, utilizing neural networks to learn the region of attraction associated with a given controller. Other learning-based approaches for Lyapunov function determination include \cite{seeger2004gaussian, berkenkamp2016safe, berkenkamp2017safe, chow2018lyapunov}. In \cite{chang2019neural}, the focus extends to simultaneously learning both the control and the Lyapunov function, providing provable guarantees of stability in larger regions of attraction. This approach directly addresses non-polynomial continuous dynamical systems, does not presume the availability of control functions other than an initial state, and employs generic feed-forward network representations without manual design. In order to train a Lyapunov neural network, unlike conventional back propagation there are no true labels for the output values of the network. The work in \cite{chang2019neural} proposes the "Lyapunov Risk" loss function as
%

\begin{equation}
\begin{gathered}
L^\text{Lyapunov}(\theta) = \\
\mathop \mathbb{E} \limits_{x \sim \mathcal{D}} \left[ \max{(0, -V_\theta(x)) + \max{(0, \nabla_{f_u}V_\theta(x))} + V_\theta^2(x^*)} \right].
\end{gathered}
\label{eq:lyapunov_loss}
\end{equation}

This loss function may be employed for the training of a neural network using empirical data, facilitating the acquisition of a stabilizing Lyapunov function. Consequently, the derived Lyapunov function can be applied in conjunction with sampling-based methods to obtain the \ac{RoA} of the controller.

\subsection{System Dynamics}

\begin{figure*}[t]
\centering
\subfloat[]{\includegraphics[width=0.45\linewidth]{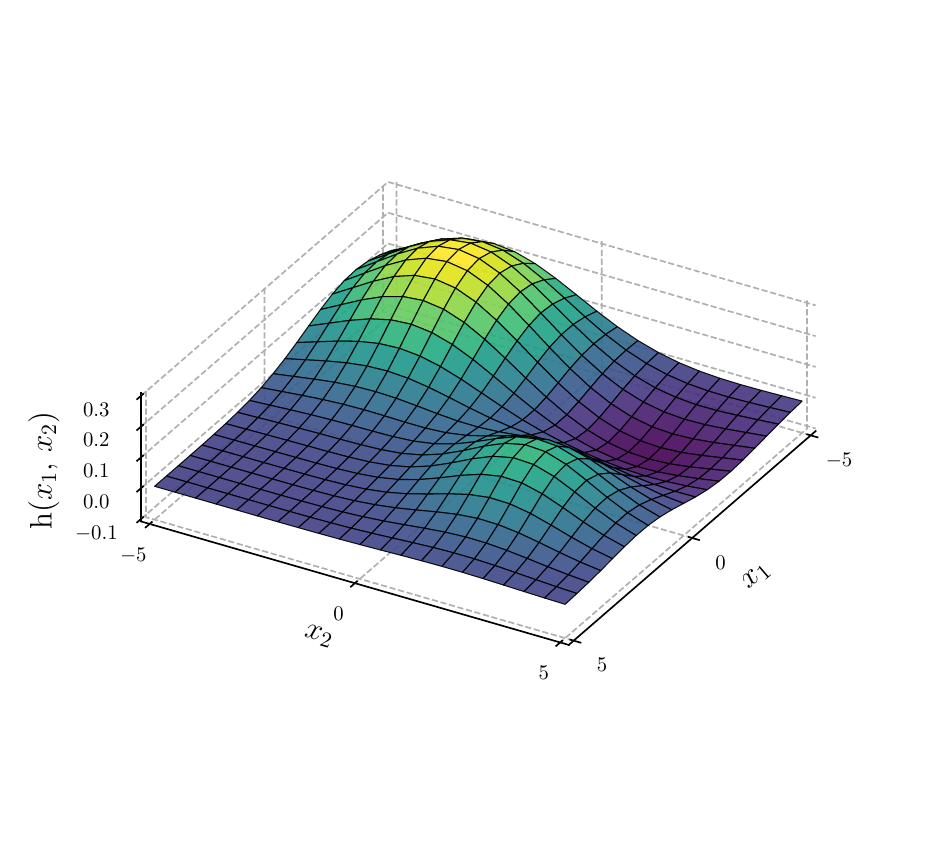}}
\subfloat[]{\includegraphics[width=0.35\linewidth]{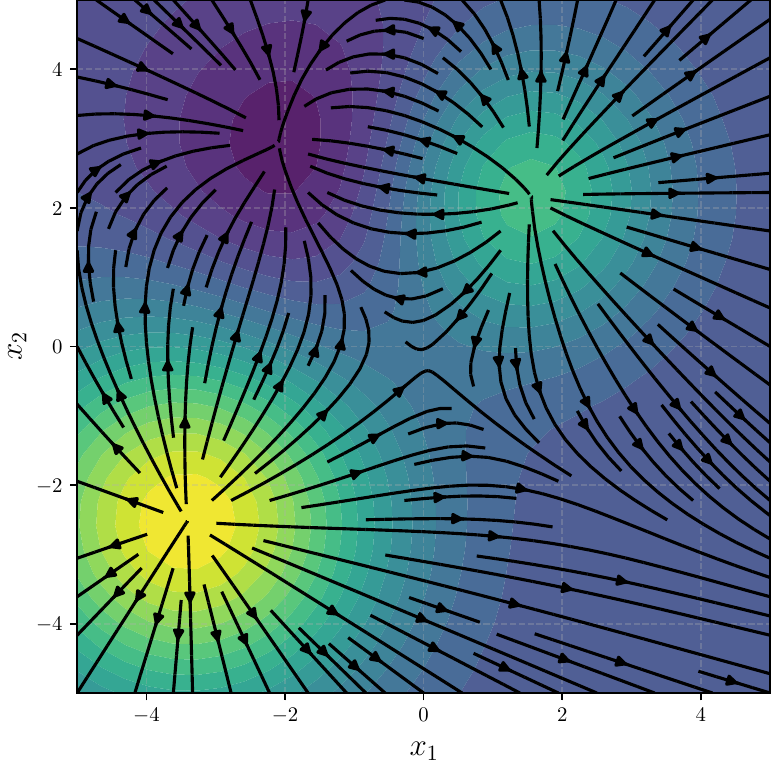}}
\caption{Dynamics used for simulating the proposed controller. The potential function $h$ is represented on the left side and its gradient that describes the vector field $f$ is depicted on the right side for the range $x_1 \in [-5, 5]$, $x_2 \in [-5, 5]$.}
\label{fig:system}
\end{figure*}

Simulation experiments on a two-dimensional toy problem have proven very useful for understanding the dynamics of the algorithm. The control problem is formulated as a two-dimensional first-order system of the form 
\begin{equation}
\begin{gathered}
  {{\dot x}_1} = {f_1}({x_1},{x_2}) + {u_1} \hfill \\
  {{\dot x}_2} = {f_2}({x_1},{x_2}) + {u_2} \hfill,  
\end{gathered}
\label{eq:system}
\end{equation}
where $x=[x_1, x_2]^T$ is the state vector. The function $f=[f_1, f_2]^T$ is obtained by taking the gradient of a potential function $h(x_1, x_2)$ as
\begin{equation}
{f_i}({x_1},{x_2}) = {\nabla _{{x_i}}}h({x_1},{x_2}) = \frac{{\partial h({x_1},{x_2})}}{{\partial {x_i}}}. 
\end{equation}
The potential function $h$ is defined as a sum of Gaussians in the form
\begin{equation}
h({x_1},{x_2}) = \sum\limits_{i = 1}^3 {{\alpha _i}\exp \left( {\frac{{{{\left( {{x_1} - {\mu _{i1}}} \right)}^2} + {{\left( {{x_2} - {\mu _{i2}}} \right)}^2}}}{{2{\sigma _i}^2}}} \right)}, 
\label{eq:h}
\end{equation}
with parameters given in \reftab{tab:gaussianparams}. The choice of dynamics $f$ is arbitrary and was chosen to keep the dimension and complexity of the system low to avoid the curse of dimensionality inherent to \ac{RL}. The resulting vector field $f$ and potential function $h$ are shown in \reffig{fig:system}. The reward function during training is defined in the form
\begin{equation}
R(x, u) = - 0.01 e_x^T e_x - 0.01 u^T u
\end{equation}
with $e_x = x - x^*$ and $u = [u_1, u_2]^T$. The control inputs are saturated to verify $u_i \in [-0.5, 0.5]$ and the states are normalized from the original range $[-5, 5]^2$ to $[-1, 1]^2$. 

\begin{table}[t]
\centering
\caption{Parameters of the potential function $h$ described in \refeq{eq:h}}
\label{tab:gaussianparams}
\begin{tabular}{@{}rrrrr@{}}
\toprule
$i$ & $\alpha_i$ & $\mu_1i$ & $\mu_1i$ & $\sigma_i$\\
\midrule
1 & -0.1 & -1.4 & 2.5 & 1.8 \\
2 & 0.2 & 1.3 & 2.2 & 1.5 \\
3 & 0.3 & -3.4 & -2.5 & 2 \\
\bottomrule
\end{tabular}
\end{table}

\section{Sequential RL controller with Neural Lyapunov certificate}
\label{sec:controller}

In this study, we introduce two sequential control methodologies. The initial approach, denoted as the \ac{TPC}, employs a hierarchical structure of interconnected controllers to guide the system from an initial state to a predefined goal state. Within this framework, each controller possesses a distinct equilibrium state and \ac{RoA}, represented as nodes in a tree structure. The system progresses sequentially through these nodes until it reaches the desired goal state, treating the intermediate states as milestones or sub-goals. However, a limitation of the \ac{TPC} arises when there is a change in the goal state, rendering the previously learned controller nodes obsolete. Consequently, the design process requires repetition, incurring computational expenses. Hence, the \ac{TPC} can be regarded as a "point-to-point" controller, and its adaptability to changes in the goal state may pose challenges.

To overcome this constraint, we also introduce an alternative approach known as the \ac{GPC}. In this graph-based method, local controllers populate the entire valued state space, irrespective of the initial and goal states. When a target goal state is selected, only a single controller at that specific point requires training. The graph of controllers can efficiently formulate a trajectory connecting any initial condition to the goal state, thereby meeting the control requirements. As a result, the \ac{GPC} is characterized as a "space-to-space" controller, offering advantages in adaptability compared to the "point-to-point" nature of the \ac{TPC}.

In this section, we establish the foundational aspects of controller design for a specified goal state, whether intermediate or final. The controller design process remains consistent regardless of the chosen approach (\ac{TPC} or \ac{GPC}). After identifying a suitable point as a node within the network, the initial step involves designing and training a controller based on \ac{RL}. This controller aims to achieve local stabilization in the vicinity of the selected node. For the implementation and evaluation of the proposed algorithms in this study, we utilize the Stable-Baselines3 library (version 2.1.0) \cite{stable-baselines3}, a widely employed reinforcement learning library. The \ac{PPO} controller's hyperparameters employed in this study are detailed in \reftab{tab:ppo_hyperparameters}.

\begin{table}[t]
\centering
\caption{PPO Algorithm Hyperparameters}
\label{tab:ppo_hyperparameters}
\begin{tabular}{@{}ll@{}}
\toprule
\textbf{Hyperparameter} & \textbf{Value} \\
\midrule
Policy network architecture & n - 128 - 64 - m \\
Value network architecture & n - 128 - 64 - 1 \\
Activation function & LeakyReLU \\
Learning rate & 3$\times$10\textsuperscript{-4} \\
Discount Factor ($\gamma$) & 0.99 \\
Number of Epochs & 10 \\
Batch Size & 256 \\
Clip Parameter ($\epsilon$) & 0.2 \\
Value Function Coefficient & 0.5 \\
Number of parallel environments & 8 \\
Train timesteps & 10\textsuperscript{6} \\
\bottomrule
\end{tabular}
\end{table}

In the training phase of the controller, initial values are deliberately chosen at random within an interval proximate to the selected equilibrium of the controller. This intentional selection is aimed at ensuring the controller's robust learning capability in stabilizing initial conditions within its local vicinity. This step is crucial, setting the foundation for the subsequent computation of quantitative attraction certificates.

After the controller, represented by the \ac{RL} agent, undergoes training, it becomes crucial to identify a region within which the controller is effective and can reliably stabilize any initial condition. Therefore, determining its \ac{RoA} becomes essential. In this study, the Lyapunov stability criterion is employed to assess the stability of the designed controller. Neural networks are trained to synthesize Control Lyapunov functions, acting as a form of function approximation. The inputs to this network include the $n$-dimensional states of the dynamic system, and the network outputs the computed Lyapunov value for the given state ($V = V_\theta(x)$). By leveraging the previously defined Lyapunov Risk Loss function, this network can be effectively trained. Note that the Lie derivative term of the Lyapunov risk is approximated by 
\begin{equation}
\nabla_{f_u} V_\theta(x) \approx \sum\limits_{i = 1}^n \frac{V_\theta(x_i + h) - V_\theta(x_i)}{h} \frac{{d{x_i}}}{{dt}}, 
\end{equation}
the parameter $h=10^{-3}$ is introduced as a minute constant for the purpose of approximating numerical derivatives. Throughout the training process, we uniformly sample 10\textsuperscript{3} initial conditions in close proximity to the controller's goal, thereby acquiring a training set comprising state-space points. The deep learning model employed for the Lyapunov function in this study is implemented using PyTorch (version 2.0.1) \cite{paszke2019pytorch}. The hyperparameters of the network are detailed in \reftab{tab:lyapunov_hyperparameters}.

\begin{table}[t]
\centering
\caption{Lyapunov Neural Network Hyperparameters}
\label{tab:lyapunov_hyperparameters}
\begin{tabular}{@{}ll@{}}
\toprule
\textbf{Hyperparameter} & \textbf{Value} \\
\midrule
Network architecture & n - 100 - 100 - 1 \\
Activation Functions & ReLU - ReLU - GELU \\
Optimizer & Adam \\
Learning Rate & 10\textsuperscript{-2} \\
Batch Size & 32 \\
Epochs & 50 \\
\bottomrule
\end{tabular}
\end{table}

After obtaining the Lyapunov function, it can be utilized to compute the \ac{RoA}. The \ac{RoA} estimate is represented as a ball around each controller, characterized by a specific radius $\eta$. To determine the maximum $\eta$ for each controller, 10\textsuperscript{3} uniformly distributed initial states are sampled in increasing proximity (up to an upper bound $\eta_{\text{ub}}$) to the equilibrium. These states are then evaluated using the Lyapunov network, and the Lyapunov stability criterion (\refeq{eq:lyapunov_criterion}) is verified. If all sampled points satisfy the criteria, the sampling radius is incrementally increased. This iterative process continues until counterexamples are identified in the sampled points or until an upper bound is reached. The steps involved in obtaining individual controllers for each selected node center are summarized in \refalg{alg:controller}.

\begin{algorithm}[H]
\caption{Reinforcement Learning-Based Controller Synthesis for Individual Nodes}
 \label{alg:controller}
\begin{algorithmic}[1]
\Require ppo\_params, lyapunov\_params, $\eta_{\text{ub}}$, $\delta \eta$
\Function{NodeController}{$x_k$}
	\State env $\gets$ MDP with $x_k$ as target state $x^*$
	\State $\pi_{\theta} \gets \text{PPO}(\text{ppo\_params})$
	\State Train $\pi_{\theta}$  
	\State $X_{\text{train}} \gets \text{sample } 10^4 \text{ points near } x_k$ 
	\State $V_{\theta} \gets \text{NN}(\text{lyapunov\_params})$
	\State Train $V_{\theta}$ with loss function \refeq{eq:lyapunov_loss} 
	\State $\eta \gets 0$  
	\While{$\eta < \eta_{\text{ub}}$} 
		\State $\eta \gets \eta + \delta \eta$ 
		\State $X_{\text{test}} \gets \text{sample } 10^3 \text{ in dist. } [\eta, \eta + \delta \eta]$ from $x_k$ 
		\State Lyapunov\_criteria $\gets$ evaluate \refeq{eq:lyapunov_criterion} on  $X_{\text{test}}$ with $V_{\theta}$
		\If{Lyapunov\_criteria not satisfied}
			\State break
		\EndIf
	\EndWhile
	\State \textbf{return $\pi_{\theta k}$, $\eta_k$ }
\EndFunction
\end{algorithmic}
\end{algorithm}

\section{\acl{TPC}}
\label{sec:tree}

This section provides an overview of the foundations of the tree-based algorithm. The controller leverages the PPO controller and the previously outlined stabilization method to construct a tree of interconnected controllers, facilitating reaching a specified goal from a given initial state. Following the introduction of the algorithm's structure, the subsequent presentation and discussion will focus on experimental results conducted on a two-dimensional dynamic system, shedding light on its effectiveness and performance.

\subsection{Design of \acl{TPC}}

The proposed algorithm begins by formulating a controller at the goal state $x_\text{goal}$. Following a methodology akin to randomized planning algorithms, the algorithm constructs a controller tree $\mathbf{T}$ through random sampling within a defined region of the state space $\mathcal{R}_\text{bound}$. The existing tree of controllers is then expanded towards a randomly selected sample point $x_k$. Ensuring the integrity of the chosen node is crucial; it must lie within the union of all preceding controller \ac{RoA}s, denoted as $\mathcal{R}_k$. This precaution is taken to ensure that, after achieving stability at the node center, there exists a node capable of progressing the trajectory toward the desired goal state.

Another condition is imposed to prevent the chosen center point for a node from being in close proximity to preceding node centers. This condition is essential to guarantee the creation of sparse controllers that can effectively cover the state space, thereby aiding in the identification of a suitable trajectory. Let $\mathcal{C}_k$ denote the sets of all previous node centers. A randomly sampled point $x_k$ is deemed an acceptable candidate for the next node center only if it meets the condition
\begin{equation}
\begin{gathered}
\left( x_k \in \mathcal{R}_\text{bound} \right) \, \land \, \left( \exists x_i \in \mathcal{C}_k, \, {\left\| {x_i - x_k} \right\|_2} \geqslant \rho \right) \\
\, \land \, \left( \forall x_\text{obstacle} \in \mathcal{O}, \, {\left\| {x_k - x_\text{obstacle}} \right\|_2} \geqslant \alpha \right),
\end{gathered}
\label{eq:pcond}
\end{equation}
where $\eta_{ub} / 2< \rho < \eta_{lb}$ is selected such that the controllers construct a sparse tree. The final constraint \ref{eq:pcond} is relevant in situations where obstacle avoidance is a desired objective. In such instances, the candidate point must adhere to the condition that it remains at a distance no less than the clearance parameter $\alpha > [ \eta_{ub} / 2 + r_{obs} ]$ from the center of all obstacles denoted as $x_\text{obstacle}$.

Upon adding a new "node" to the tree, we learn a stabilizing controller specifically for that node. The \ac{RoA} for this controller is estimated using the methodologies discussed earlier. It is crucial to confirm that the computed \ac{RoA} surpasses the specified lower bound $\eta_\text{lb}$. This verification ensures that the formulated controller can effectively cover a significant region within the state space. If the \ac{RoA} falls short of this criterion, the controller is rejected, and no further points are sampled in the vicinity of the previously invalidated point.

In this phase, we determine the parent node for the selected point. Among all nearby nodes that encompass the point $x_k$ within their \ac{RoA}, we identify the node closest to the point and set it as the parent node $i$. These computations are performed backward, initiating from the goal and extending into the state space until reaching the starting state. As a result, the backward tree transforms into a complex network of local controllers with the capability to attract initial conditions and guide them toward the goal, all while providing formal certificates of stability for the nonlinear system. The algorithm concludes when the starting state is enclosed within the constructed trees region of attraction $\mathcal{R}_k$. The steps involved in designing the \ac{TPC} are summarized in \refalg{alg:tpc}.

\begin{algorithm}[H]
\caption{Sequential Control Algorithm Based on a Tree of Connected Controllers (\ac{TPC})}
\label{alg:tpc}
\begin{algorithmic}[1]
 \Require $x_\text{start}, x_\text{goal}, \eta_{\text{lb}}$
  \State $\pi_{\theta g}$, $\eta_g \gets $ \Call{NodeController}{$x_\text{goal}$} 
  \State $\mathbf{T} \gets \{ (x_\text{goal}, \pi_{\theta g}, \eta_g, \text{NULL}) \}$ 
  \For{$k=1$ to $K$} 
  	\State $x_k \gets \text{random sample}$ 
  	\If{$x_k$ satisfies \refeq{eq:pcond}}
  		\State $\pi_{\theta k}$, $\eta_k \gets $ \Call{NodeController}{$x_k$}
  		\If{$\eta_k > \eta_{lb}$}
  			\State i $\gets$ pointer to parent node in $\mathbf{T}$ containing $x_k$
 			\State $\mathbf{T} \gets \mathbf{T} \cup \{ (x_k, \pi_{\theta k}, \eta_k, i) \}$ 
 			
		\EndIf
 	\EndIf
	\If{$x_\text{start} \in \mathcal{R}_{k}$} 
		\State break
	\EndIf
 \EndFor
\end{algorithmic}
\end{algorithm}

\subsection{Result of \acl{TPC}}

\begin{figure}[t]
\centering
\subfloat[]{\includegraphics[width=0.4\textwidth]{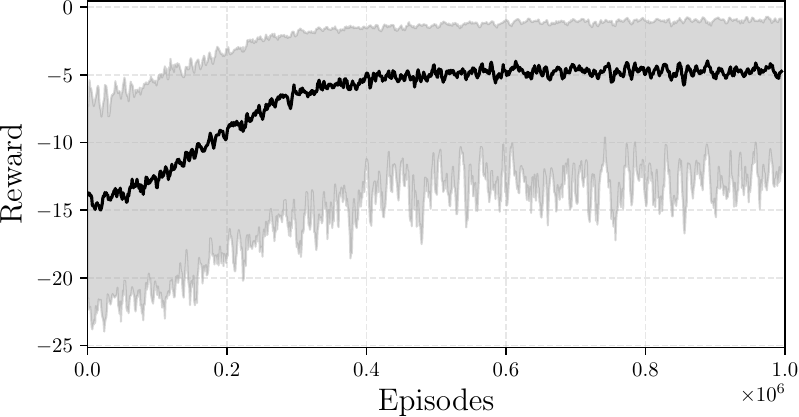}\label{fig:tr}} \\
\subfloat[]{\includegraphics[width=0.4\textwidth]{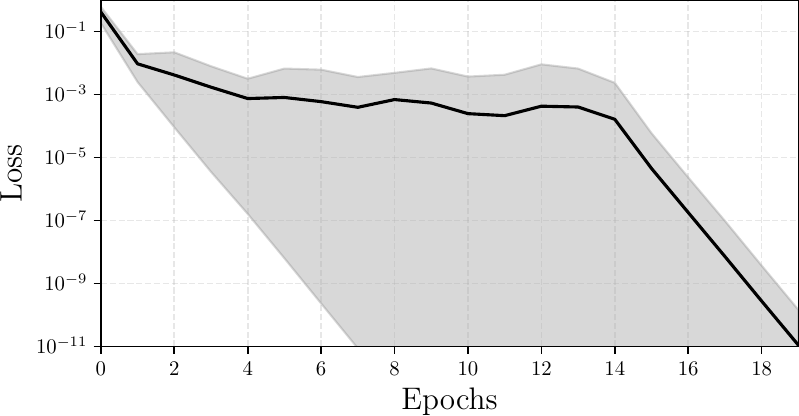}\label{fig:tv}} \\
\caption{(a) Average cumulative episode reward observed during the training of the RL agent for the \acl{TPC}. (b) Lyapunov Loss trends throughout the training of the Lyapunov neural networks for computing the region of attraction for each RL controller. The black line represents the average among the 23 controllers, while the shaded region illustrates the maximum and minimum bounds of the computed Rewards and Loss respectively.}
\label{fig:trv}
\end{figure}

In the initial simulation of the proposed algorithm, we tested the tree-structured controller. The parameters used were $x_\text{goal} = [4, 4]^T$ and $x_\text{start} = [-4, -4]^T$. The task assigned to the \ac{TPC} controller was to compute a tree of sequential controllers aimed at reaching the goal state from the initial starting point. The completion of this task resulted in a tree comprising a total of 23 nodes. During the training of all node controllers (\ac{PPO} agents), the resulting reward values are depicted in \reffig{fig:tr}. These values represent an average across all computed nodes of the tree. Notably, the rewards show a gradual saturation at -5, starting from an initial reward of -15. Following the computation of controllers, a Lyapunov neural network was trained for each individual node. This network is intended for use in calculating the \ac{RoA} for each node. The loss over 20 epochs of training the network is visualized in \reffig{fig:tv}.

\begin{figure}[t]
\centering
\subfloat[]{\includegraphics[width=0.45\linewidth]{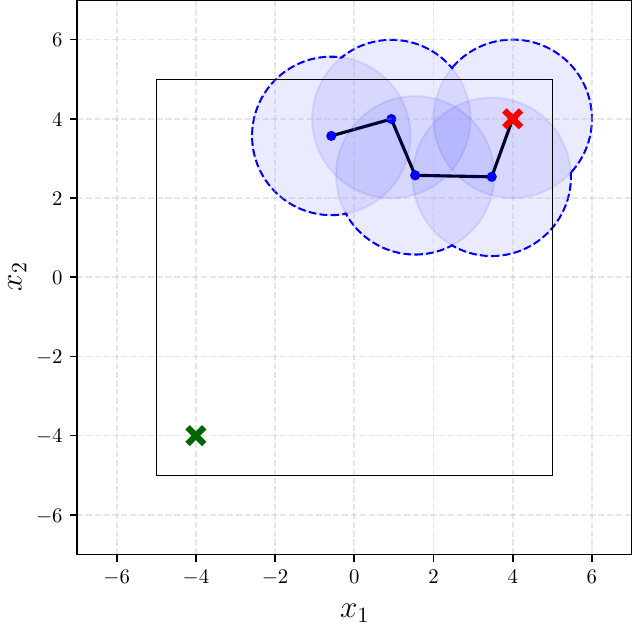}\label{fig:t1}} \qquad
\subfloat[]{\includegraphics[width=0.45\linewidth]{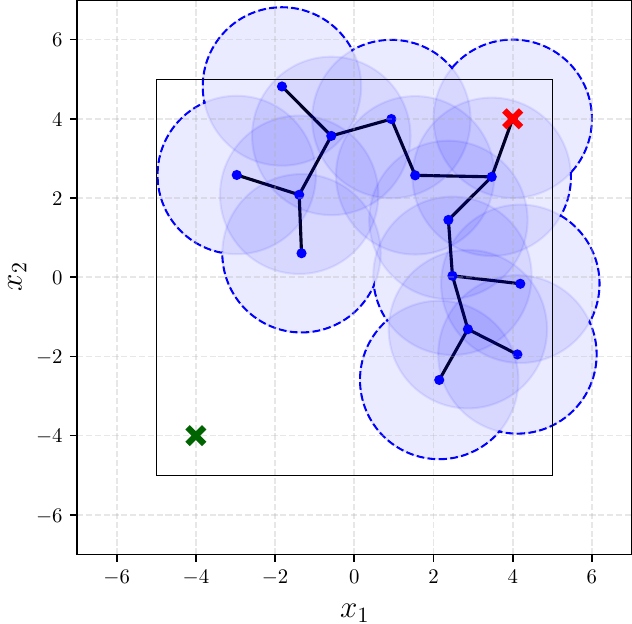}\label{fig:t2}} \\
\subfloat[]{\includegraphics[width=0.45\linewidth]{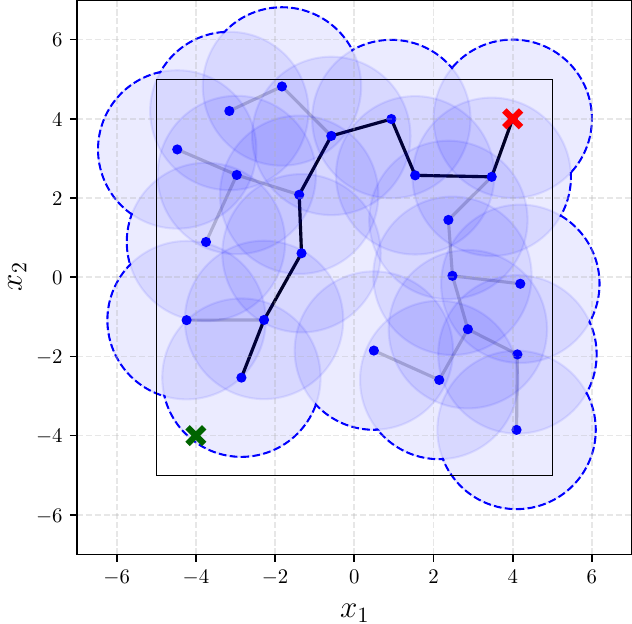}\label{fig:tf}} \qquad
\subfloat[]{\includegraphics[width=0.45\linewidth]{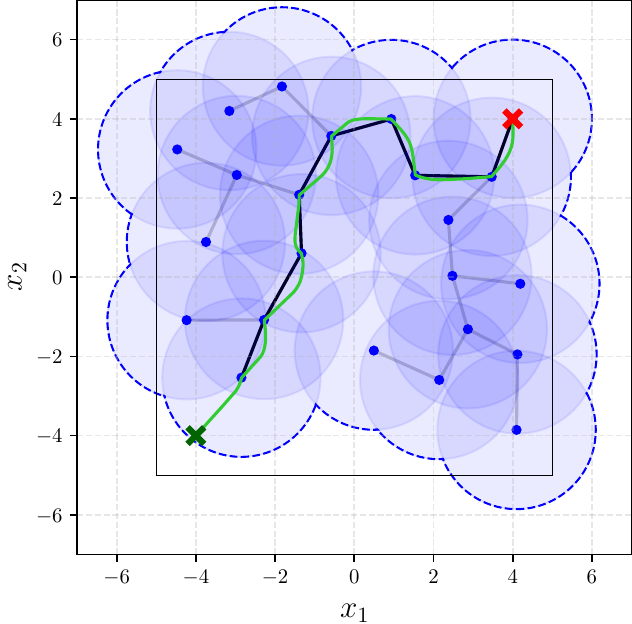}\label{fig:tt}}
\caption{The \acl{TPC} applied to the dynamic system defined in \refeq{eq:system}. (a) Snapshot of the algorithm with 5 nodes. (b) Snapshot with 15 nodes. (c) Initial state reaching the covered space of the tree at 23 nodes, with irrelevant branches pruned. (d) Dynamic system trajectory following the constructed controller tree. Red cross: goal state $x_\text{goal} = [4, 4]^T$; Green cross: starting state $x_\text{start} = [-4, -4]^T$; Black square: bounded region $\mathcal{R}_\text{bound}$; Blue dots: controller node center locations $x_k$; Blue shaded region: individual node regions of attraction; Blue dashed shaded region: overall region of attraction of the tree ($\mathcal{R}_k$); Black lines: connections (edges) between tree nodes; Green line: system trajectory $x(t)$.}
\label{fig:t}
\end{figure}

The dynamics of the proposed tree-structured algorithm are depicted in \reffig{fig:t}. The algorithm initiates at the goal state, formulating and stabilizing local controllers while branching the tree within the bounded state space region $\mathcal{R}_\text{bound} = [-5, 5]^2$. With each step, the search region expands until it encompasses the desired initial state. Snapshots of the algorithm in action at 5 and 15 tree nodes are presented in \reffig{fig:t1} and \reffig{fig:t2}, respectively. By the time the tree comprises 23 nodes, the entire area covered by the tree $\mathcal{R}_k$ includes the starting state. In \reffig{fig:tf}, the path connecting the final node to the first calculated node is highlighted, and all irrelevant branches are "pruned," resulting in a path featuring 10 nodes. The system is capable of traversing a trajectory from the nodes in sequence until reaching the globally desired equilibrium at $x_\text{goal}$. Controller switching occurs when the system approaches the equilibrium of a controller node, specifically when $\left| {x(t) - x_k} \right|_2 < 0.01$. Finally, the trajectory of the system following the nodes in the tree is illustrated in \reffig{fig:tt}.

\begin{figure}[t]
\centering
\subfloat[]{\includegraphics[width=0.45\linewidth]{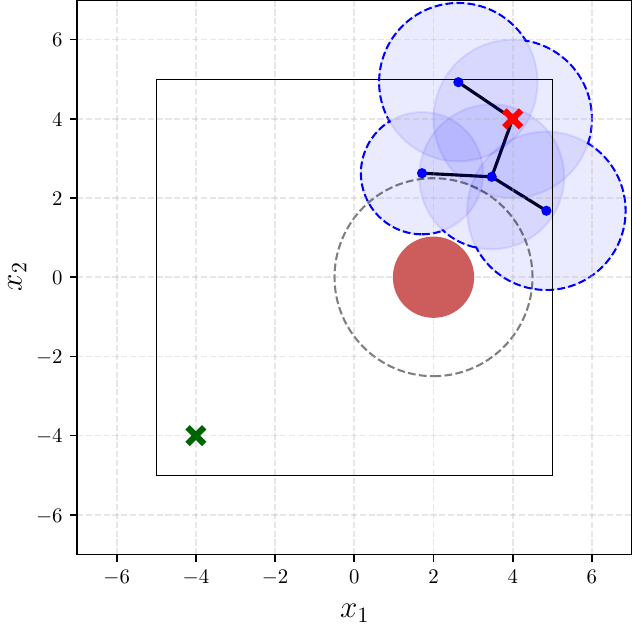}\label{fig:to1}} \qquad
\subfloat[]{\includegraphics[width=0.45\linewidth]{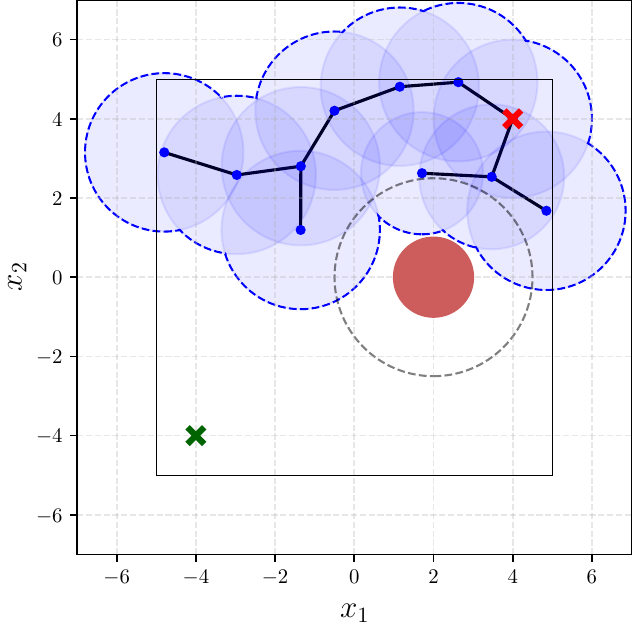}\label{fig:to2}} \\
\subfloat[]{\includegraphics[width=0.45\linewidth]{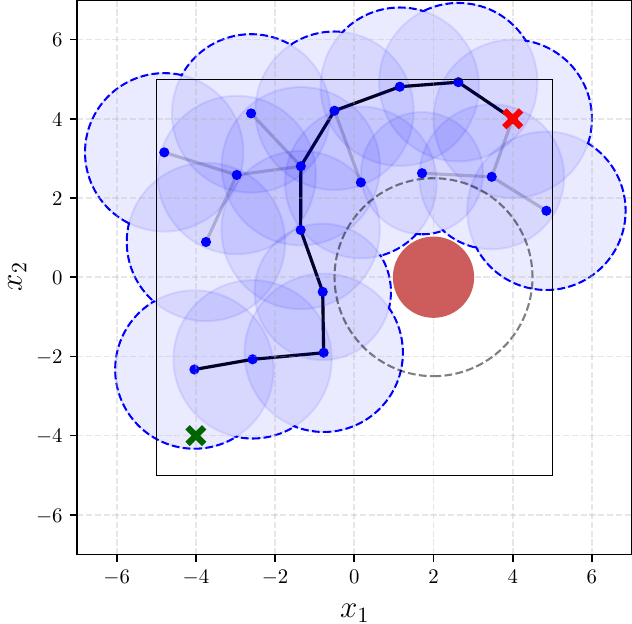}\label{fig:tof}} \qquad
\subfloat[]{\includegraphics[width=0.45\linewidth]{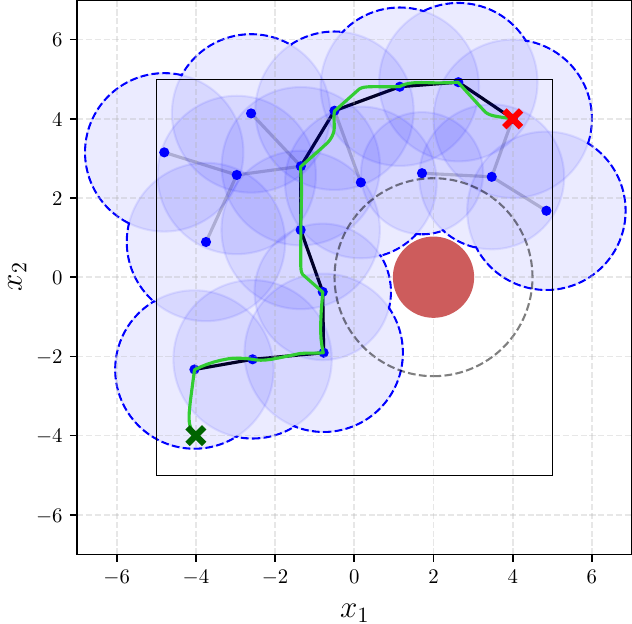}\label{fig:tot}}
\caption{The \acl{TPC} applied to the dynamic system defined in \refeq{eq:system} with obstacle with radius 1 at $x_\text{obstacle} = [2, 0]^T$. (a) Snapshot of the algorithm with 5 nodes. (b) Snapshot with 11 nodes. (c) Initial state reaching the covered space of the tree at 18 nodes, with irrelevant branches pruned. (d) Dynamic system trajectory following the constructed controller tree. Red cross: goal state $x_\text{goal} = [4, 4]^T$; Green cross: starting state $x_\text{start} = [-4, -4]^T$; Black square: bounded region $\mathcal{R}_\text{bound}$; Red circle: obstacle; Black dashed circle: obstacle clearance; Blue dots: controller node center locations $x_k$; Blue shaded region: individual node regions of attraction; Blue dashed shaded region: overall region of attraction of the tree ($\mathcal{R}_k$); Black lines: connections (edges) between tree nodes; Green line: system trajectory $x(t)$.}
\label{fig:to}
\end{figure}

To showcase the algorithm's obstacle avoidance capability, an additional simulation is conducted with an obstacle present in the state space, as depicted in \reffig{fig:to}. The obstacle, centered at $x_{obstacle} = [2, 0]^T$ with a radius of 1, is taken into consideration. A clearance of $\alpha = 2.5$ is set to ensure the avoidance of node selection in close proximity to the obstacle. Upon simulating the \ac{TPC} algorithm in this obstacle-laden scenario, a tree consisting of 18 nodes is generated. Following the pruning of irrelevant branches, the final path between the start and goal states comprises 10 nodes, as illustrated in \reffig{fig:tof}. The system trajectory utilizing the controller is visualized in \reffig{fig:tot}.

\section{\acl{GPC}}
\label{sec:graph}

This section provides an overview of the foundations of the graph-based algorithm. The controller leverages the PPO controller and the previously outlined stabilization method to construct a graph of interconnected controllers, facilitating goals from any initial states in the region of interest. Following the introduction of the algorithm's structure, the subsequent presentation and discussion will focus on experimental results conducted on a two-dimensional dynamic system, shedding light on its effectiveness and performance.

\subsection{Design of \acl{GPC}}

This approach involves constructing a connected graph of controllers $\mathbf{G}$ within a delimited region $\mathcal{R}_\text{bound}$ in the state space. Once this interconnected graph is established, it facilitates the generation of trajectories from any given initial condition to a desired goal state. Similar to the \ac{TPC} method, this technique entails the random selection of points within the state space and the design of controllers at these locations. The criteria for accepting a sampled point $x_k$ as a new node candidate for the graph are identical to the constraints outlined in the tree-based algorithm \refeq{eq:pcond}. The main departure from the previous algorithm lies in the initialization of the graph. In contrast to the earlier approach where the graph started at the goal state and expanded towards the initial state, the current method evolves to create a sparse network of connected controllers spanning the specified bounded region of the state space, regardless of the desired initial and goal states. The algorithm concludes when no valid $x_k$ can be identified, signaling that the graph effectively covers the entire designated region ($\mathcal{R}_{bound} \subseteq \mathcal{R}_{k}$).

After the creation of individual controllers, only one controller needs to be generated at the desired goal state. Subsequently, a connected graph is established, where each node is linked to all controller nodes containing the node within their \ac{RoA}, referred to as the valid nodes set $\mathbf{V}$. It is essential to note that the edges are directional, reflecting the variation in $\eta$ size within their respective \ac{RoA}. In brief, if a controller center is within another controller's \ac{RoA}, the reverse may not necessarily be true. To navigate the graph from any initial state to the goal state controller, shortest path finding algorithms can be employed. In this work, Dijkstra's algorithm \cite{DIJKSTRA1959} is utilized to determine the optimal path between controller nodes in the final graph. The steps involved in designing the \ac{GPC} are succinctly outlined in \refalg{alg:gpc}.

\begin{algorithm}[H]
\caption{Sequential Control Algorithm Based on a Graph of Connected Controllers (\ac{GPC})}
\label{alg:gpc}
\begin{algorithmic}[1]
\Require $x_\text{start}, x_\text{goal}, \eta_{\text{lb}}, \mathcal{R}_{bound}$
\For{$k=1$ to $K$} 
 	\State $x_k \gets \text{random sample}$ 
 	\If{$x_k$ satisfies \refeq{eq:pcond}}
 		\State $\pi_{\theta k}$, $\eta_k \gets $ \Call{NodeController}{$x_k$}
 		\If{$\eta_k > \eta_{lb}$}
			\State $\mathbf{G} \gets \mathbf{G} \cup \{ (x_k, \pi_{\theta k}, \eta_k) \}$ 
			
	\EndIf
	\EndIf
	\If{$\mathcal{R}_{bound} \subseteq \mathcal{R}_{k}$} 
		\State break
	\EndIf
\EndFor
\State $\mathbf{G} \gets \mathbf{G} \cup \{ (x_\text{goal}, \pi_{\theta g}, \eta_g) \}$ 
\State $\mathbf{V}$ $\gets$ pointers to valid nodes in $\mathbf{G}$ containing each $x_k$
\State Trajectory $\gets$ \Call{Dijkstra}{$x_\text{start}, x_\text{goal}, \mathbf{G}, \mathbf{V}$} 
\end{algorithmic}
\end{algorithm}

\subsection{Result of \acl{GPC}}

\begin{figure}[t]
\centering
\subfloat[]{\includegraphics[width=0.4\textwidth]{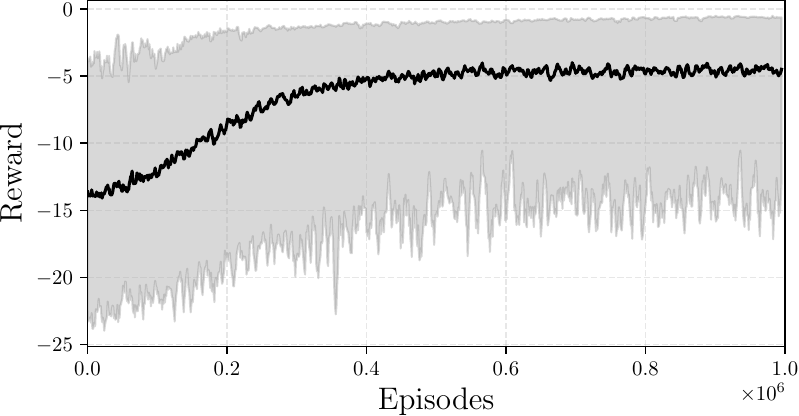}\label{fig:gr}} \\
\subfloat[]{\includegraphics[width=0.4\textwidth]{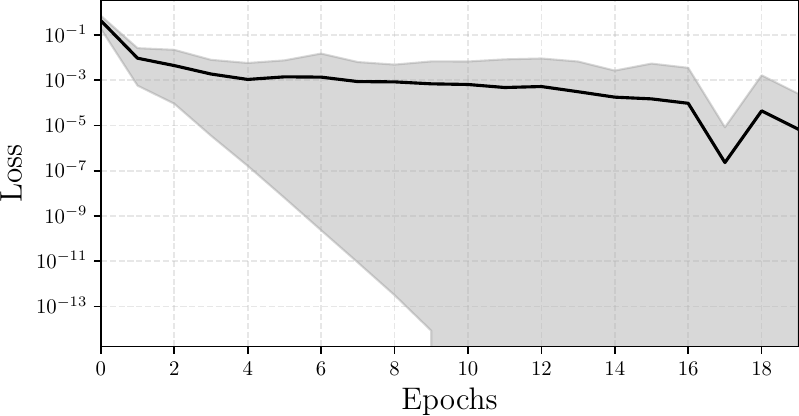}\label{fig:gv}} \\
\caption{(a) Average cumulative episode reward observed during the training of the RL agent for the \acl{GPC}. (b) Lyapunov Loss trends throughout the training of the Lyapunov neural networks for computing the region of attraction for each RL controller. The black line represents the average among the 37 controllers, while the shaded region illustrates the maximum and minimum bounds of the computed Rewards and Loss respectively.}
\label{fig:grv}
\end{figure}

The results of simulating the proposed \ac{GPC} controller are presented in this section. As previously stated, this controller is "space-to-space", so there is no need of determining the starting and goal states in the initial stage of controller design. Individual nodes are added in succession, spanning the entirety of the bounded region after 37 nodes are added. During the training of all node controllers (\ac{PPO} agents), the resulting reward values are depicted in \reffig{fig:gr}. These values represent an average across all computed nodes of the tree. Similar to the previous tree based approach, the rewards again show a gradual saturation at -5, starting from an initial reward of -15. The Lyapunov network loss over 20 epochs of training the network is visualized in \reffig{fig:gv}.

\begin{figure}[t]
\centering
\subfloat[]{\includegraphics[width=0.45\linewidth]{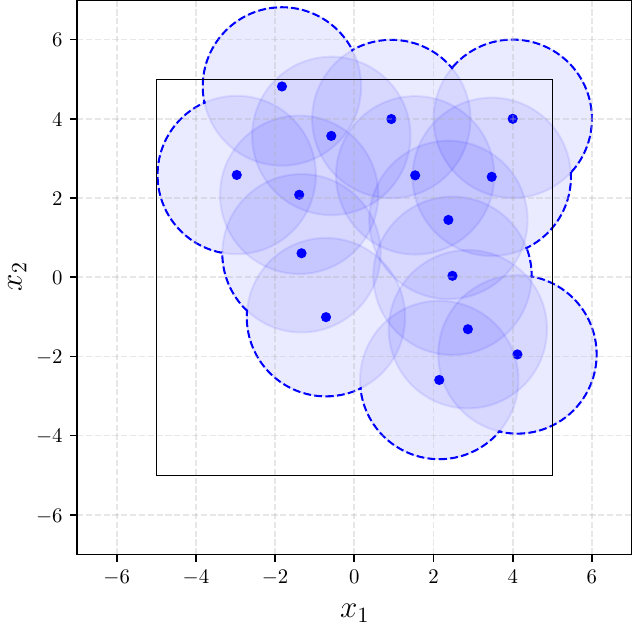}\label{fig:g1}} \qquad
\subfloat[]{\includegraphics[width=0.45\linewidth]{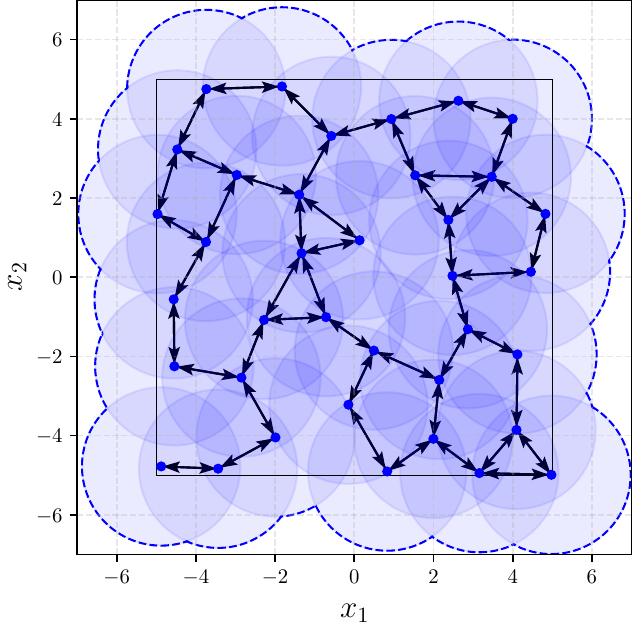}\label{fig:g2}} \\
\subfloat[]{\includegraphics[width=0.45\linewidth]{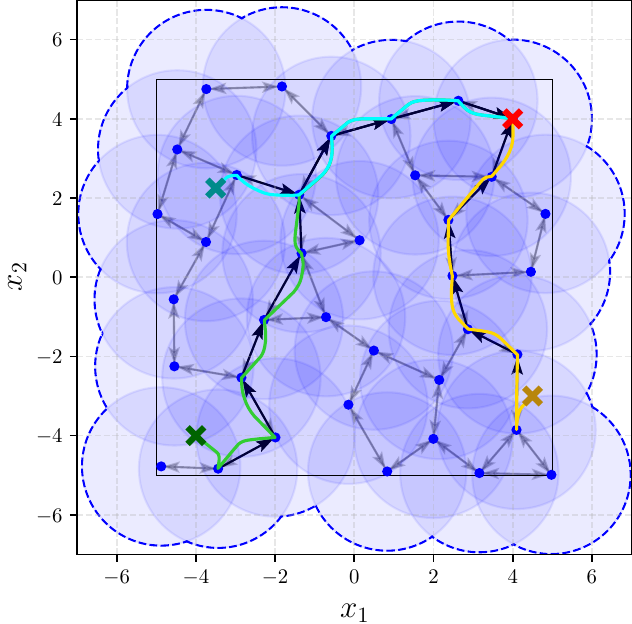}\label{fig:gf1}} \qquad
\subfloat[]{\includegraphics[width=0.45\linewidth]{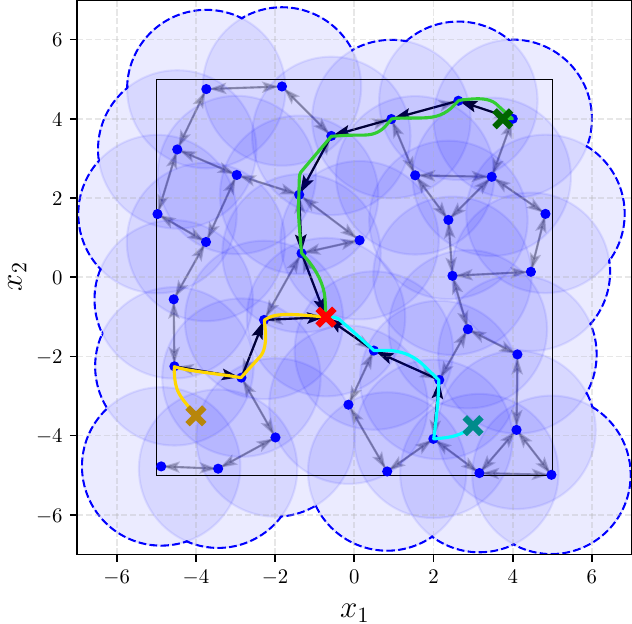}\label{fig:gf2}}
\caption{The \acl{GPC} applied to the dynamic system defined in \refeq{eq:system}. (a) Snapshot of the algorithm with 15 nodes. (b) Final graph with the total 37 interconnected nodes and edges. (c) Experiment with $x_\text{goal}^{(1)}=[4, 4]^T$ and three initial states at $x_\text{start}^{(1a)} = [-4, -4]^T$, $x_\text{start}^{(1b)} = [-3.5, 2.25]^T$, and $x_\text{start}^{(1c)} = [4.5, -3]^T$. The sequence of controllers reaching the goal is found using Dijkstra algorithm and is highlighted. (d) Experiment with $x_\text{goal}^{(2)}=[-0.75, -1]^T$ and three initial states at $x_\text{start}^{(2a)} = [3.75, 4]^T$, $x_\text{start}^{(2b)} = [3, -3.75]^T$, and $x_\text{start}^{(2c)} = [-4, -3.5]^T$. Red cross: goal state $x_\text{goal}$; Green/Yellow/Cyan cross: starting states $x_\text{start}^i$; Black square: bounded region $\mathcal{R}_\text{bound}$; Blue dots: controller node center locations $x_k$; Blue shaded region: individual node regions of attraction; Blue dashed shaded region: overall region of attraction of the graph ($\mathcal{R}_k$); Black lines: connections (edges) between graph nodes; Green/Yellow/Cyan line: system trajectories $x^i(t)$.}
\label{fig:gf}
\end{figure}

The algorithm's operation is illustrated in \reffig{fig:gf}. Initially, it spans the space, as depicted in \reffig{fig:g1}, where the algorithm is shown with 15 nodes. After the design reaches 37 nodes, filling the entire region of interest, this phase of the algorithm concludes. Subsequently, the graph is generated by connecting all possible adjacent nodes reachable from each node, as depicted in \reffig{fig:g2}. Notably, due to discrepancies in the sizes of \ac{RoA} for different nodes, not all connections are bidirectional. This becomes more apparent in a later example involving the presence of an obstacle. After constructing the \ac{GPC}, the first example involves selecting a goal state at the point $x_\text{goal}^{(1)}=[4, 4]^T$ and three initial states at $x_\text{start}^{(1a)} = [-4, -4]^T$, $x_\text{start}^{(1b)} = [-3.5, 2.25]^T$, and $x_\text{start}^{(1c)} = [4.5, -3]^T$. Once each initial condition's starting node is determined, the Dijkstra algorithm calculates the shortest path from that node to the desired goal node. The dynamic system then follows the computed path between sequential controllers, similar to the \ac{TPC} method (\reffig{fig:gf1}).

To showcase the algorithm's ability to adapt to changes in goal or initial states, the states are modified to $x_\text{goal}^{(2)}=[-0.75, -1]^T$, and three new initial states are chosen: $x_\text{start}^{(2a)} = [3.75, 4]^T$, $x_\text{start}^{(2b)} = [3, -3.75]^T$, and $x_\text{start}^{(2c)} = [-4, -3.5]^T$, as illustrated in \reffig{fig:gf2}. 

%

\begin{figure}[t]
\centering
\subfloat[]{\includegraphics[width=0.45\linewidth]{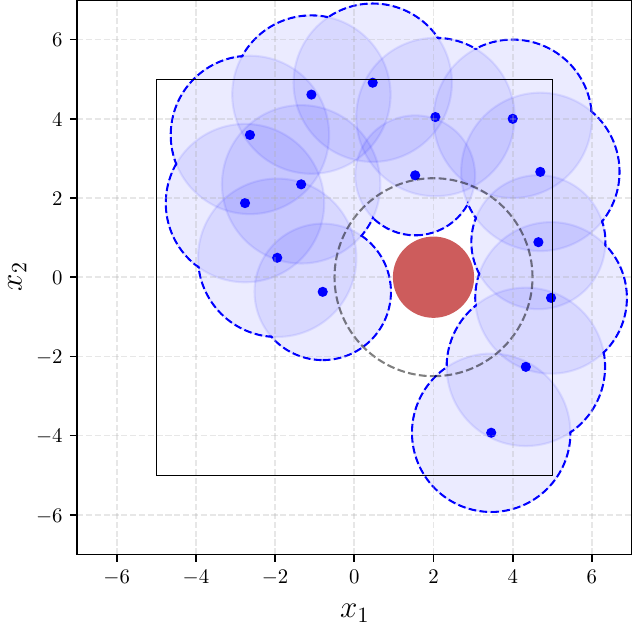}\label{fig:go1}} \qquad
\subfloat[]{\includegraphics[width=0.45\linewidth]{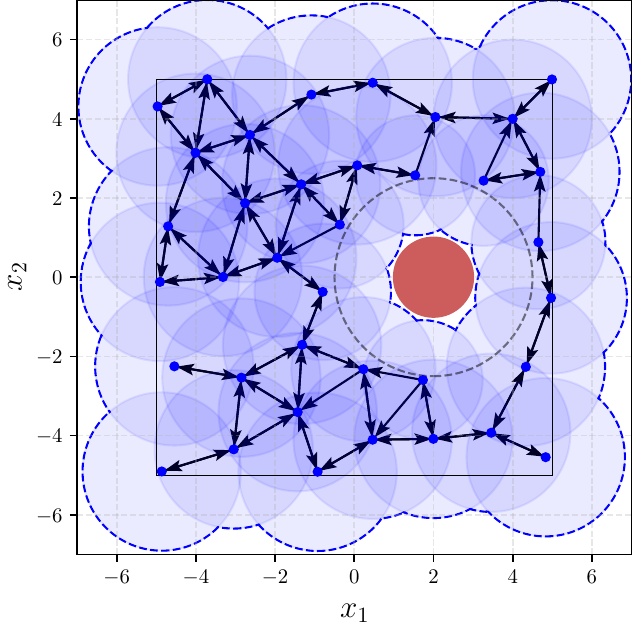}\label{fig:go2}} \\
\subfloat[]{\includegraphics[width=0.45\linewidth]{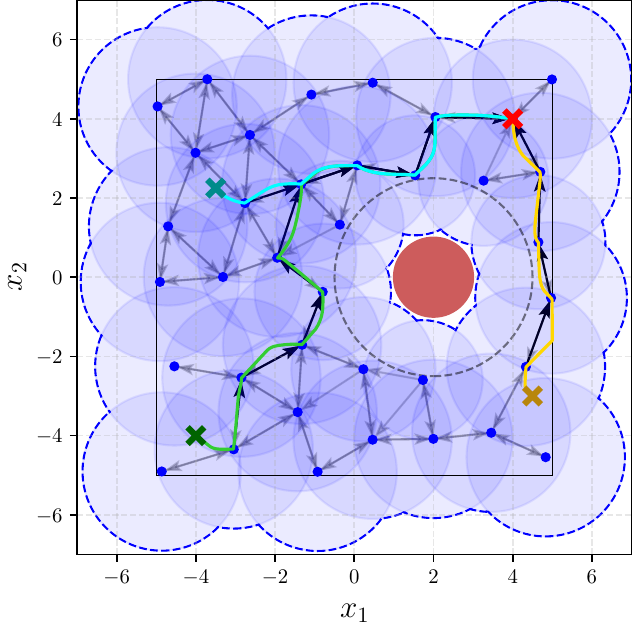}\label{fig:gof1}} \qquad
\subfloat[]{\includegraphics[width=0.45\linewidth]{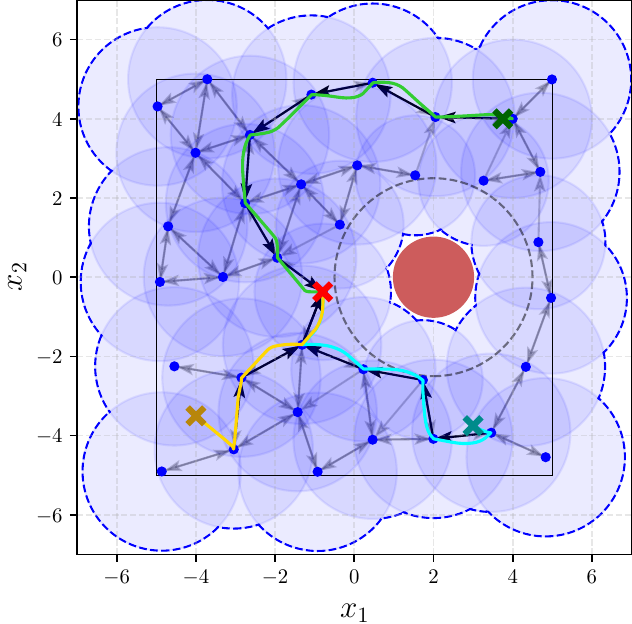}\label{fig:gof2}}
\caption{The \acl{GPC} applied to the dynamic system defined in \refeq{eq:system} with obstacle with radius 1 at $x_\text{obstacle} = [2, 0]^T$. . (a) Snapshot of the algorithm with 15 nodes. (b) Final graph with the total 37 interconnected nodes and edges. (c) Experiment with $x_\text{goal}^{(1)}=[4, 4]^T$ and three initial states at $x_\text{start}^{(1a)} = [-4, -4]^T$, $x_\text{start}^{(1b)} = [-3.5, 2.25]^T$, and $x_\text{start}^{(1c)} = [4.5, -3]^T$. The sequence of controllers reaching the goal is found using Dijkstra algorithm and is highlighted. (d) Experiment with $x_\text{goal}^{(2)}=[-0.75, -1]^T$ and three initial states at $x_\text{start}^{(2a)} = [3.75, 4]^T$, $x_\text{start}^{(2b)} = [3, -3.75]^T$, and $x_\text{start}^{(2c)} = [-4, -3.5]^T$. Red cross: goal state $x_\text{goal}$; Red circle: obstacle; Black dashed circle: obstacle clearance; Green/Yellow/Cyan cross: starting states $x_\text{start}^i$; Black square: bounded region $\mathcal{R}_\text{bound}$; Blue dots: controller node center locations $x_k$; Blue shaded region: individual node regions of attraction; Blue dashed shaded region: overall region of attraction of the graph ($\mathcal{R}_k$); Black lines: connections (edges) between graph nodes; Green/Yellow/Cyan line: system trajectories $x^i(t)$.}
\label{fig:gof}
\end{figure}

To showcase the capabilities of the \ac{GPC} controller in environments with obstacles, a test is conducted similar to the previous section. An obstacle is introduced at $x_{obstacle} = [2, 0]^T$ with a radius of 1, and the clearance parameter is set to $\alpha = 2.5$. The completed graph, after adding 37 nodes, is depicted in \reffig{fig:go2}. Notably, many connections between nodes are unidirectional, but a guarantee exists that there is a path between every pair of nodes in the graph. The system trajectory, navigating from three initial states at $x_\text{start}^{(1a)} = [-4, -4]^T$, $x_\text{start}^{(1b)} = [-3.5, 2.25]^T$, and $x_\text{start}^{(1c)} = [4.5, -3]^T$ to the goal state $x_\text{goal}^{(1)}=[4, 4]^T$, is illustrated in \reffig{fig:gof1}. Similarly, the trajectory from $x_\text{start}^{(2a)} = [3.75, 4]^T$, $x_\text{start}^{(2b)} = [3, -3.75]^T$, and $x_\text{start}^{(2c)} = [-4, -3.5]^T$ to $x_\text{goal}^{(2)}=[-0.75, -1]^T$ is shown in \reffig{fig:gof2}.

\section{Tree-Structured vs. Graph-Structured Controller}
\label{sec:discussion}

A comprehensive comparison of the two proposed methods in this study is outlined in \reftab{tab:comparison}. Addressing the question of "When should each method be used?" is pivotal. The primary distinction between the two methods lies in the inherent structure of the controllers and the interconnection of nodes. The \ac{TPC} adopts a Tree structure, while the \ac{GPC} employs a graph structure. Both controllers consist of nodes with PPO controllers, and the design, training, and stabilization of these controllers follow a similar approach. As previously discussed, the key divergence in the learning process between the controllers is the criteria for initiation and termination. In the case of \ac{TPC}, the process initiates by designing node controllers from the goal state, branching out into the bounded state space until encompassing the start state. On the other hand, \ac{GPC} adds nodes to the graph regardless of the goal and initial states. Its objective is to span the entire desired region with locally stable controllers. Once the graph is constructed, a trajectory can be defined, traversing sequentially through the controllers for any desired start and goal state. Nodes in the tree structure exhibit a parent-child relationship, where each node (excluding the root node or goal) has precisely one parent. In contrast, the directed graph of \ac{GPC} implies that each edge has a direction, and connections between nodes can be bidirectional. Furthermore, the graph is cyclic, containing cycles, in contrast to the acyclic nature of the tree structure. Both methods choose node positions in a strategic manner outlined before as to have the capability for obstacle avoidance. 

The primary distinction lies in the computational cost associated with each controller. As evidenced by the results, the tree controller is capable of identifying a valid solution with fewer nodes. In the scenario without an obstacle, the tree controller synthesized only 23 nodes, as opposed to the graph, which required 37 nodes. Likewise, when an obstacle is present, the tree controller synthesized 18 nodes compared to the graph's 37 nodes. This underscores the advantage of the \ac{TPC} in rapidly finding solutions when a single valid solution is crucial. Conversely, the \ac{GPC} excels in situations where the initial state may vary or remain unknown at the onset of calculations. As mentioned earlier, once a tree is constructed, the tree controller can only manage the initially defined state or a limited set of initial conditions falling within the tree's region of attraction. In contrast, the graph, designed with nodes spanning the entire region of interest, enables the controller to stabilize the system for any chosen initial condition. A similar scenario arises when altering the goal state. The tree controller necessitates a complete reconstruction from scratch, whereas the graph controller simply adds a single node for the new goal state and connects it to the existing graph.

A crucial aspect to consider is the sequence of controller switching. Trees exhibit a hierarchical structure, featuring a designated root node (goal state) that serves as the starting point for accessing all other nodes. Given that each node has only one parent, the sequence of nodes leading from the initial state to the goal state is clearly defined. In contrast, graphs lack this inherent hierarchy, often possessing multiple paths between any two nodes. To address this complexity, Dijkstra's algorithm was employed to determine the optimal node sequence between two states in the graph.

\begin{table*}[t]
  \centering
  \caption{Comparison of \acl{TPC} and \acl{GPC}}
  \label{tab:comparison}
  \begin{tabular}{@{}lll@{}}
    \toprule
    \textbf{Attribute} & \textbf{\acl{TPC}} & \textbf{\acl{GPC}} \\
    \midrule
    Controller structure & Tree & Graph \\
    Node controllers & PPO & PPO \\
    Directionality & No & Bidirectional \\
    Obstacle avoidance & Yes & Yes \\
    Upfront cost & \textbf{Low} & High \\
    Initial-Goal states & (1 initial state) to (1 goal state) & \textbf{(All initial states) to (1 goal state)} \\
    Cost of changing goal state & Whole tree must be reconstructed & \textbf{Only 1 node must be added to the graph} \\
    Cost of changing initial state & Whole tree must be reconstructed & \textbf{None} \\
    Pathfinding Algorithm & None & Dijkstra \\
    \bottomrule
  \end{tabular}
\end{table*}

\section{Conclusion}
\label{sec:conclusion}

This paper introduces a novel sequential control algorithm employing localized \ac{PPO} agents as nodes. To assess the\ac{RoA}, Lyapunov stability criteria are employed, with a neural network trained at each node to determine a locally valid Lyapunov function. Notably, this approach eliminates the necessity for explicit dynamics models, relying instead on data-driven techniques for acquiring the node controller and stabilization. The paper proposes two distinct algorithms for connecting node controllers to establish a sequential controller. Firstly, a tree-structured method is presented for achieving "point-to-point" control, facilitating motion control from an initial state to a desired goal state. Additionally, a graph-structured approach is outlined for scenarios requiring "space-to-space" control. While the latter method incurs an upfront computational cost, it offers the advantage of reusing the structured graph in cases of changes in the desired goal state or initial conditions. Both algorithms are equipped to handle obstacle avoidance scenarios. This capability ensures that the algorithms can identify a tree (or graph) of interconnected controllers that navigate around obstacles safely. The effectiveness of the proposed controllers is demonstrated through application to a two-dimensional first-order system. Results indicate the algorithm's ability to design a sequence of connected Reinforcement Learning \ac{RL}-based controllers, effectively addressing the control problem. However, certain areas warrant further exploration. Future research avenues include the incorporation of path-finding algorithms to optimize node selection, particularly in higher-dimensional systems. Additionally, it is crucial to evaluate the controller's performance through experimentation on a test platform. This would allow for a comprehensive examination of the control framework in real-world operational scenarios, considering interactions and noise dynamics between the robotic system and its environment.

\bibliographystyle{IEEEtran}
\bibliography{references}

\newpage

%
%
%
%

\vfill

\end{document}